# Super-resolution measurements related to uncertainty relations in optical and biological fluorescence systems

Y.Ben-Aryeh

Physics Department, Technnion-Israel Institute of Technology, Haifa, 32000, Israel

e-mail: phr65yb@ph.technion.ac.il

FAX: 972-4-8295755

*keywords:*

super-resolution

optical high spatial modes

biological fluorescence high-resolution

saturation effects

coherent interference

analogy to lasers

ABSTRACT

Super-resolution effects in optical and fluorescence biological systems are analyzed and their relations with uncertainty relations are discussed. Super-resolutions obtained in the optical systems, including especially NSOM, SIM and hyperlens, are related to an increase of the spatial frequencies in the object plane leading to very small effective wavelengths and thus the resolution is increased far beyond the Abbe limit. Super-resolution measurements obtained in the fluorescent biological systems STED, FPALM and RESOLFT are treated. An example of a four-level STED system is analyzed in analogy to a four-level laser system, but the special space dependence of the STED light is taken into account, restricting the fluorescence from extremely small volume, and thus extremely high resolution is obtained. Localization of individual molecules is described by the FPALM method where interference between coherent fluorescent photons is taken into account. While in the STED method very high laser intensities are needed in its variant method known as RESOLFT the super-resolution measurements can be obtained by much weaker light intensities. This new method is analyzed and the reasons for a such large reduction in light intensities are explained.

## 1. Introduction

The observation of sub-wavelength structures with microscopes is restricted usually by the *Abbe diffraction limit* [1], by which light with a wavelength $\lambda = \lambda_0/n$ travelling in a medium with a refractive index $n$ and angle $\theta$ will make a spot with radius

$$d = \frac{\lambda_0}{2(n \sin \theta)} \quad . \tag{1}$$

The term $n \sin \theta$ appearing in the denominator is called the numerical aperture (NA) and the Abbe limit for ordinary microscopes is of order $\lambda/2$.

To increase the resolution one may use UV and X-ray microscopes which increase the resolution due to their shorter wavelength. Such microscopes suffer from lack of contrast in biological systems, are expensive and may also damage the sample. One can increase the resolution also by using electron microscopes where the electron mass wavelength is relatively small but electrons can interact with the biological systems and then would lead to distorted images.

Recently, very high resolutions have been obtained in biological systems which are much better than the Abbe limit by using *far field* optical methods, including super-resolution *fluorescence microscopy*. The amount of articles on this topic is enormous and I refer to some review articles [2-7] for the experimental details of such high resolution experiments. While the far field optical methods have been verified and analyzed in many works I would like, in the present work, to discuss the relations between super-resolution measurements and uncertainty relations.

We follow in the present article the idea that super-resolution effects can be related to space-momentum uncertainty relations. In the one dimensional $x$ direction the *classical* uncertainty relation can be given as $\Delta k_x \Delta x \geq 1/2$, where $\Delta k_x$ and $\Delta x$ are the uncertainties in the wavevector $k_x$ and the space coordinate $x$, respectively. (By multiplying the two sides of this equation by Planck constant $\hbar$, one gets the Heisenberg quantum uncertainty relation $\Delta p_x \Delta x \geq \hbar/2$, but for our purpose it is enough to use the classical uncertainty).

In the general case, the super-resolution measurements can be related to unceretainties in a three-dimensional space. Usually the super-resolution effects are described in the lateral $(x, y)$ space but super-resolution in the axial $z$ direction is also important and can be realized by longitudinal optical focusing techniques.

I describe, shortly, in the next Section the optical super-resolution methods known as Near Field Scanning Optical Microscopes ( NSOM) [8-13], hyperlens system [14,15], and structured illumination microscopy (SIM) [2,4]. I show that the common feature of these methods is their relation to the above uncertainty relation. The idea behind these methods is that by using patterns with high *spatial frequencies* the 'effective' wavelength becomes smaller and thus increases the resolution. There are also other optical methods which are based on using high spatial frequencies e.g. using multi-photon processes with a large frequency uncertainty . One should notice that $c\Delta k = \Delta(2\pi c/\lambda) = \Delta\omega$, so that uncertainties in the wavevector $\vec{k}$ involves a wide distribution for $\omega$, and vice versa. One should notice therefore that in the super-resolution measurements the increase in the spatial frequencies should involve a wide distribution in the frequency $\omega$. Such effect is opposite to that of lasers where there $\Delta\omega$ tends to be very small while $\Delta t = \Delta x/c$ becomes very large.

While the optical methods lead to a resolution which is better than the Abbe limit , it is quite surprising to find that the resolution obtained in *far field* microscopic biological measurements is even much better. Most of the super-resolution measurements in biological systems are based on *fluoresence* light emission. A main problem in using fluorescence emission for super-resolution measurements is that the emission from different biological structures will overlap and thus would lead to destructive interference. Therefore, different methods have been developed to localize the fluorescence emission from individual biological micro-elements, (see e.g. [16-28]).

A special type of fluorescence scanning for super-resolution measurements which is known as Stimulated Emission Depletion (STED) microscopy has been developed by S.W. Hell and his colleagues [16-23]. The size of the fluorescence spot ,i.e., the point spread function (PSF) , can be reduced by employing a second *STED laser* in addition to the first *excitation laser*, where by stimulated emission it prevents the fluorescent emission in the outer region of the excitation spot. The STED laser beam has a wavelength longer than the detected fluorescence, therefore, the stimulated emission will also emit at longer wavelength not seen by the detector. The shape of the effective fluorescence spot which determines the spatial resolution is sharpened by the second STED beam laser via the stimulated emission. Treating a desired depletion beam profile at the focal point is one of the major challenges to improve the resolution. My point is that in the STED method one gets a very rapid space dependence of the number of molecules in the excited states from which fluorescence is obtained so that the Fourier transform of such space dependence would lead to high spatial frequencies reducing the 'effective' wavelength by some orders of magnitudes and thus leads to the super-resolution effects.

There is a fundamental problem in a quantitative treatment of stimulated emission in biological super-resolution measurements which is quite often ignored. Stimulated radiation leads both to emission and absorption where their ratio depends only on the number of molecules in the upper and lower levels, respectively. Taking into account this effect and certain theoretical methods from the field of lasers I analyze in Section 3 an example of the STED method.

In another method known as Fluorescence Photoactivated Localization Microscope (FPALM) [24-28] localization of individual molecules is treated . I explain in Section 3, the FPALM method in relation to uncertainty relations and show that in a single molecule fluorescence localization the 'effective' wavelength can be reduced by a coherent interference between $N$ fluorescenting photons reducing the effective wavelength as $\lambda/\sqrt{N}$. This effect is related in Section 3 to other lasers theories.

STED can also work if there are two different molecular states of a photoswitchable probe. This scheme employs a fluorescence probe that can operate reversibly between a fluorescent and dark state [23]. This more general approach is termed Reversible Saturable Optical Fluorescence Transition (RESOLFT) which applies to all ensemble techniques based on a stimulated emission between any two molecular states. In the ordinary STED method very strong EM fields are needed but in a variant of this method known as RESOLFT the effective spot size $\Delta x$ becomes very small even by using relatively much weaker EM fields. I will analyze in Section 3 the RESOLFT method and discuss the reasons why in this method one can obtain super-resolution measurements by EM fields which are *weaker by many order of magnitudes* relative to those used in the original STED method.

Although there are other methods for obtaining super-resolution measuremets, the differences between them are mainly in the optical techniques for getting smaller effective wavelength and their relations with the uncertainty relations are similar to those described in the present article. I use an approach for super-resolution in biological systems in Section 3 which is consistent with the analyses given for optical systems in Section 2. I demonstrate the present approach by analyzing the above special systems.

## 2. Super-resolution effects related to Helmholtz equation and to high spatial frequencies in optical systems

In homogeneous medium the Helmholtz equation is given as

$$(nk_0)^2 = k_x^2 + k_y^2 + k_z^2 \qquad , \qquad (2)$$

where $k_0 = (2\pi)/\lambda_0$, $\lambda_0$ is the wavelength in vacuum (or approximately in air), $n$ is the index of refraction (assumed to be approximately real), and $k_x, k_y, k_z$ are the wavevector components.

The assumption of scalar EM fields simplifies very much the analysis of super-resolution effects, and the main properties of diffraction effects can be related to such fields [29,30]. In order to overcome the Abbe diffraction limit various methods have been developed. I describe here, shortly, physical systems in which super-rsolution measurements in optical systems have been obtained which are far beyond the Abbe limit. Using the same line of thinking, I will analyze in the next Section the present approach for super-resolution measurements in biological systems:

The NSOM method is based on the use of evanescent waves obtaining the relation

$$(k_x^2 + k_y^2) > (nk_0)^2 \qquad , \qquad (3)$$

where $k_z$ becomes imaginary, i.e., there is a decay of the wave in the $z$ direction. The increase of the components of the wavevector $\vec{k}$ in the $(x,y)$ plane decreases the "effective" value of the wavelength in this plane, and thus increases the resolution. But the evanescent wave decay, however, in the $z$ direction perpendicular to the object plane, so that in order to "capture" the fine structures which are available in the evanescent waves we need to put detectors very near to the object plane from which the EM waves are propagating. Different super-resolution effects which are obtained by the use of evanescent waves have been analyzed in previous articles [9-13].

Propagating waves with high wave vectors have been realized also by the use of the hyperlens [14,15] composed of certain *meta materials*. The principles by which such system is obtained can be explained as follows. The dispersion of EM waves in cylindrical coordinates for an *isotropic medium* is
given by

$$k_r^2 + k_\theta^2 = \varepsilon \frac{\omega^2}{c^2} \qquad (4)$$

where $k_r$ and $k_\theta$ are the radial and tangential wave vector components, $\varepsilon$ is the permittivity of the isotropic medium, $\omega$ is angular frequency and $c$ is the speed of light. The angular momentum number $m$ satisfies the relation

$$m = k_\theta r \qquad , \tag{5}$$

where $r$ is the distance from the center. The tangential component $k_\theta$ of the wave vector for a certain $m$ number increases towards the center as $k_\theta \propto 1/r$, and then according to Eq. (4), below a certain ciritical $r$ value $k_r$ becomes imaginary so that the angular states are converted to evanescent waves.

In the case of *uniaxial anisotropy* the dielectric permittivity is characterized by two values: $\varepsilon_\parallel$ along the optical axis of the crystal, and $\varepsilon_\perp$ transverse to the optical axis. Propagating modes can be decomposed into two polarization states: the *ordinary* (TE) and *extraordinary* (TM) waves. For ordinary (TE) waves, the electric field vector is transverse to the optical axis and produces the same dielectric response (given by $\varepsilon_\perp$) independent of wave propagation direction. For extraordinary (TM) wave the electric field has components both along and transverse to the optical axis and the elliptic dispersion relation is given by

$$\frac{k_\perp^2}{\varepsilon_\parallel} + \frac{k_\parallel^2}{\varepsilon_\perp} = \frac{\omega^2}{c^2} \qquad , \tag{6}$$

where $k_\perp$ and $k_\parallel$ refer to wave vector components normal or parallel to the optical axis.

In the case of strong anisotropy where $\varepsilon_\perp$ and $\varepsilon_\parallel$ are of opposite signs, the dispersion relation becomes *hyperbolic* :

$$\frac{k_r^2}{\varepsilon_\theta} - \frac{k_\theta^2}{|\varepsilon_r|} = \frac{\omega^2}{c^2} \qquad , \tag{7}$$

where $\varepsilon_\theta > 0$ , $\varepsilon_r < 0$ . As the tangential component of the wave vector increases towards the center, the radial component also increases, and we get propagating waves with very large wave vectors leading to super resolution effects. The relation (7) can be obtained by certain optical techniques using such *meta materials*, and by which the hyperlens has been realized (see e.g. [14,15]).

The use of *meta materials* for obtaining super-resolution has been analyzed also for another system known as Pendry lens [31], but the transformation of evanescent waves into propagating waves is very crucial for effective realization of such system [32].

The basic concept of Structured illumination Microscopy (SIM) [2,4] is to produce interference between the unknown structure which have low spatial frequencies with another *known pattern*. If the known pattern have higher spatial frequencies the technique will present a better spatial resolution. The idea that the convolution between two spatial patterns would lead to super-resolution effects and also to other interesting phenomena has been described in a previous work [12].

## 3. Super-resolution measurements in fluorescence biological systems related to lasers theories

My argument is that there are close relations between the STED method and laser theories [30]. For demonstrating a laser theory approach to this system, I analyze here an example of four–level STED microscope described in [16,22]. In this scheme we have two molecular electronic levels where the level with smaller electronic energy is denoted as $S_0$, and that with the higher electronic energy as $S_1$. We take into account also that each of these electronic levels has corresponding vibrational energies. The first laser (the excitation laser) excites the molecules from the ground vibrational level of the electronic state $S_0$ into an high vibrational level of the electronic state $S_1$, and from there the molecules are relaxing to the vibrational ground state of $S_1$. We denote the ground vibrational state of $S_1$ as level $2$. The second laser, i.e., the STED laser, transfers the molecules by stimulated emission from level $2$ to an high vibrational level of the electronic state $S_0$, where this level will be denoted as level $1$. The molecules are relaxing from level 1 to the vibrational ground state of $S_0$. The fluorescence light is emitted from level $2$ to the ground vibrational level of $S_0$ so that the fluorescent light is in frequencies different from the STED light and the detector that detects the fluorescent light does not detect the STED light. All the experimental details of this physical system in which super-resolution measurements have been obtained, have been described in [16,22].

One might notice that the four-level scheme described here is similar to a four-level scheme used in laser theories (see e.g. Figure 13.2.6 in [30]), but as the mechanism for obtaining super-resolution in the STED method is different, we need to insert some changes in the conventional laser analysis. We use here the same optical methods developed for lasers to analyze quantitatively the STED *fluorescence mechanism.*

If a molecule is illuminated by a mono-chromatic beam of light of frequency $\nu$, with mean photon-flux density $I$ (photons per second per unit area) the probability of stimulated emission (if the atom is in the upper level) or absorption (if the atom is in the lower level) is [30]:

$$W(\nu) = I\sigma(\nu) \quad , \tag{8}$$

where $\sigma(\nu)$ is the transition cross section for spontaneous emission given by

$$\sigma(\nu) = \frac{\lambda^2}{8\pi t_{sp}} g(\nu) \quad . \tag{9}$$

Here $t_{sp}$ is the spontaneous life time, $g(\nu)$ is the line shape with a width $\Delta \nu$ around a central resonance frequency $\nu_0$. While the equations of motion for the above four-level system have been presented [16] their explicit solutions will turn to be quite complicated. In order to relate such equations to an analysis of a STED microscope it is preferable to reduce the equations to the two-level, 1 and 2, system where the stimulated emission and absorption is operating between them, in analogous way to the two-level laser systems [30].

Let us present first the equations of motion for the number densities of molecules in levels 1 and 2 *in the absence of the STED radiation.* We simplify the equations by assuming that the effects of the *excitation laser* transferring the molecules from the ground vibrational level of $S_0$ to an high vibrational level of $S_1$ and the vibrational relaxing of the molecules to the vibrational ground state of $S_1$ can be combined as an 'effective' pumping rate $R$ into level 2. Then the rate of a change in the population number densities of level 2 and 1 are given by

$$\frac{d\tilde{n}_2(t)}{dt} = R - \frac{\tilde{n}_2(t)}{\tau_{flou}} - \frac{\tilde{n}_2(t)}{\tau_{2,1}} \quad , \tag{10}$$

$$\frac{d\tilde{n}_1(t)}{dt} = -\frac{\tilde{n}_1(t)}{\tau_{vib}} + \frac{\tilde{n}_2(t)}{\tau_{2,1}} \tag{11}$$

where the tilde over the photon numbers $\tilde{n}_1(t)$ and $\tilde{n}_2(t)$ at time $t$ denotes that these numbers densities are obtained in the absence of the STED radiation. Here $\tau_{flou}$ is the decay time for the fluorescent radiation, $\tau_{2,1}$ is the decay time from level 2 to level 1 including especially the spontaneous emission, and $\tau_{vib}$ is the vibrational decay time from the level 1 to ground vibrational level of $S_0$. As a first order approximation we neglect in the present analysis transmissions with other levels.

Under steady-state conditions :

$$\tilde{n}_2 = R\tau_2 \quad ; \quad 1/\tau_2 = 1/\tau_{flou} + 1/\tau_{2,1} \quad , \tag{12}$$

$$\tilde{n}_1 = \frac{\tilde{n}_2 \tau_{vib}}{\tau_{2,1}} \quad ; \quad \tilde{n}_2 - \tilde{n}_1 = N_0 = R\tau_2(1 - \frac{\tau_{vib}}{\tau_{2,1}}) \quad , \tag{13}$$

where $N_0$ is the steady state population in the absence of the STED laser. Assuming $\tau_{vib} \ll \tau_{2,1}$ we get the approximation $\tilde{n}_1 \ll \tilde{n}_2$.

By adding the STED light, Eqs. (10-11) are changed into:

$$\frac{dn_2(\vec{x},t)}{dt} = R - \frac{n_2(\vec{x},t)}{\tau_2} - W_{STED}(\vec{x})[n_2(\vec{x},t) - n_1(\vec{x},t)] \quad , \tag{14}$$

$$\frac{dn_1(\vec{x},t)}{dt} = -\frac{n_1(\vec{x},t)}{\tau_{vib}} + \frac{n_2(\vec{x},t)}{\tau_{2,1}} + W_{STED}(\vec{x})[n_2(\vec{x},t) - n_1(\vec{x},t)] \quad , \tag{15}$$

where $W_{STED}$ is given by Eqs.(8-9) for the STED laser, i.e., $W(\nu) \to W_{STED}(\nu)$ but the critical point in the STED method is that the photon flux density $I(\vec{x})$ and consequently $W_{STED}(\vec{x},\nu)$ are functions of the coordinates location of $\vec{x}$. Under steady-state conditions we get from Eq. (14):

$$n_2(\vec{x}) = R\tau_2 - W_{STED}(\vec{x})[n_2(\vec{x}) - n_1(\vec{x})]\tau_2 \quad , \tag{16}$$

and from Eq. (16) and (15):

$$n_1(\vec{x}) = \frac{\tau_{vib}\{R\tau_2 - W_{STED}(\vec{x})[n_2(\vec{x}) - n_1(\vec{x})]\}}{\tau_{2,1}} + \tau_{vib}W_{STED}(\vec{x})[n_2(\vec{x}) - n_1(\vec{x})] \quad . \tag{17}$$

Substracting Eq. (17) from Eq. (16) and adding together all the terms which are proportional to $W_{STED}(\vec{x})$ we get:

$$n_2(\vec{x}) - n_1(x) \equiv N(\vec{x}) = R\tau_2\left[1 - \frac{\tau_{vib}}{\tau_{2,1}}\right] - W_{STED}(\vec{x})[n_2(\vec{x}) - n_1(\vec{x})]\left\{\tau_2\left[1 - \frac{\tau_{vib}}{\tau_{2,1}}\right] + \tau_{vib}\right\} \quad , \tag{18}$$

where $N(\vec{x})$ is the difference in the population number densities under steady state conditions between levels 2 and 1 at point $\vec{x}$. We notice that according to Eq. (13) that $R\tau_2\left[1-\dfrac{\tau_{vib}}{\tau_{2,1}}\right]$ is equal to $N_0$, which represents the steady state difference in the number densities of photons in the absence of the STED laser.

We define

$$\tau_s = \left\{\tau_2\left[1-\dfrac{\tau_{vib}}{\tau_{2,1}}\right]+\tau_{vib}\right\} \quad , \tag{19}$$

and then we get

$$N(\vec{x}) = \dfrac{N_0}{1+W_{STED}(\vec{x})\tau_s} \quad ; \quad \tau_s = \dfrac{1}{I_s} \quad , \tag{20}$$

where $I_S$ is the light saturatiuon intensity. Eq. (20) shows the ratio between the difference in the number of molecules $N(\vec{x}) = n_2(\vec{x}) - n_1(\vec{x})$ in the presence of the STED laser relative to their difference $N_0$ in the absence of the STED laser. Under the condition that $\tau_{vib}$ is very small relative to other parameters we obtain $n_1(\vec{x}) \ll n_2(\vec{x})$ so that the result for $N(\vec{x})$ is approximately equal to that of $n_2(\vec{x})$ from which we get the fluorescence. The above relations depend critically on the above parameters which follow from conventional use of lasers theories.

We find according to Eq. (8) that $W_{STED}(\vec{x})$ depends on the line shape function $\sigma(\nu)$. Let us assume, for example, that we have a Lorentzian line shape function given by

$$\sigma(\nu) = \dfrac{1}{\pi}\dfrac{\Delta\nu}{(\nu-\nu_0)^2+\Delta\nu^2} \quad . \tag{21}$$

where $\Delta\nu$ is the line width and $\nu_0$ is its central frequency. Eq. (20) will be correct only under the assumption of homogeneous line broadening [30] (e.g. collisions broadening) where any absorption or emission process is spread homogenously over all the line shape function. Under the assumption of inhomogeneous broadening, where there are different groups of molecules with different absorption and emission rates, Eq. (20) should be exchanged into

$$N(\vec{x}) = \frac{N_0}{\sqrt{1+W_{STED}(\vec{x})\tau_S}} \quad ; \quad \tau_S = \frac{1}{I_S} \quad . \tag{22}$$

The change of Eq. (20) to Eq. (22) for inhomogeneous broadening can be related to the effect of "hole burning" [30,33]. Strong applied signal burn holes in inhomogeneous spectral lines producing differents saturation effects for different holes. Those spectral packets which are close to the signal frequency saturate according to Eq. (20). But for closely neighboring packets which are out of resonance the saturation is not strong as the central packet. So that by averaging over different wavepackets the saturation effect becomes weaker given approximately by Eq. (22). An approximate derivation of Eq. (22) has been given in Ref. [30] p.486, for Doppler broadening, but a similar result can be given for other cases of inhomogeneous broadening. In many biological systems, saturation effects are assumed to be given by (22), as in other cases of inhomogeneous broadening.

Although we have given a detailed analysis of the STED method for the above four-level system similar analyses can be made for other systems, e.g. the three-level system [17]. The key to achieving super-resolution is the non-linear dependence of the depleted population on the STED laser when the saturation depletion level is approached. The idea developed and verified in many experiments by Hell and his colleagues is that one can use the STED laser with a space dependence which will restrict the fluorescence of the molecules from a very small volume. The STED laser causes molecules excited at the edges of a normal diffraction limited volume to be driven to the ground state without fluorescence by illuminating them with annular laser beam at a frequency that causes stimulated emission from the excited state. Only those molecules at the null (center) of the donut-shaped STED beam remain in the excited state long enough to fluorescence, resulting in emission from a highly confined volume. Such fluorescence radiation has very small space uncertainties leading to high-resolution measurements.

In the FPALM method two lasers are used to control the number of active (potentially fluorescent) molecules. A high-energy visible laser, called the activation laser, is used to activate a small number of molecules in the sample. This number is kept small by making the activation laser intensity sufficiently weak in the focal plane of the sample. Activation of molecule occurred randomly. The activation laser is turned off and molecules that were activated are read out, using a second, typically lower energy laser, called the read out laser. Readout means detection of fluorescence from activated molecules within the illuminated area. One should take into account that when molecules are present in close proximity, localization becomes inaccurate or impossible. Separating activation and imaging allows to contract the fraction of molecules in the fluorescent state at a given time, such that the activated molecules are optically resolvable from each other and precisely localized. In one imaging process the detection is over *separated* superfluorescent molecules so that there will not be any destructive interference between them. By repeating the imaging process *other* fluorescent molecules are detected and by a combination of all these imaging processes the full image is obtained.

The precision of localization process in the FPALM is given approximately by $s/\sqrt{N}$, where s is the standard deviation of the PSF and $N$ is the number of photons detected from one fluorescent molecule before it 'bleaches', i.e., before it decays or transmitted to other non-fluorescent states. In order that the fluorescence emission of one molecule will be repeated it needs to be under the interaction of the read out laser leading to a repeating process of laser excitation and fluorescence deexcitation. In each deexcitation one fluorescent photon is emitted. The fact that *cooperative* effects between $N$ photons can lead to high resolution effects has been emphasized and explained in previous works [9,34]. One should notice that the factor $\sqrt{N}$ is the standard deviation in the number of photons in a coherent state and it is related to statistical Poisson distribution [27]. In principle, if one repeats the experiment $N$ times for a single fluorescent photon the same $s/\sqrt{N}$ dependence will be obtained. While noise effect can decrease the super-resolution effects, under the condition that these noise terms are relatively small the above $1/\sqrt{N}$ dependence will be the dominant effect.

Breaking the diffraction barrier by the RESOLFT method requires: 1) Two states A and B that are distinct in their optical properties. State A is fluorescence activated while state B is fluorescence inhibited. 2) The optical transition from state A to state B takes place at a rate $k_{Ab}(\vec{x}) = \sigma_{eff} I(\vec{x})$ that is proportional to the light intensity $I(\vec{x})$ applied at point $\vec{x}$, while the reverse transition take place at a rate $k_{BA}$ which is independent of the light intensity $I(\vec{x})$ (In principle $\sigma_{eff} = \sigma_2 - \sigma_1$ where $\sigma_2 I(\vec{x})$ leads to the stimulated transition from A to B and $\sigma_1 I(\vec{x})$ leads to the inverse stimulated emission from B to A, but we assume $\sigma_2 \gg \sigma_1$). We have the equation of motion

$$\frac{dN_A(\vec{x},t)}{dt} = -\sigma_{eff} I(\vec{x}) N_A(\vec{x},t) + k_{BA} N_B(\vec{x},t) = -\frac{dN_B(\vec{x},t)}{dt} \quad , \tag{23}$$

In Eq. (23) we have assumed that the *change in time* of the total number densities of molecules in state A plus those in state B at any point $\vec{x}$ vanishes, i.e., $N_A(\vec{x},t) + N_B(\vec{x},t)$ is preserved in time. We assume also the initial conditions at time $t = 0$

$$N_A(\vec{x},t=0) = N \quad ; \quad N_B(\vec{x},t=0) = 0 \tag{24}$$

where $N$ is independent of $x$. We find according to Eqs. (23-24) that:

$$N_A(x,t) + N_B(x,t) = N \tag{25}$$

where $N$ is independent also of $t$. By using Eqs. (24-25) we transform Eq. (23) into the form

$$\frac{dN_A(\vec{x},t)}{dt} = -\left[\sigma_{eff} I(\vec{x}) + k_{BA}\right] N_A(\vec{x},t) + k_{BA} N \tag{26}$$

The solution of Eq. (26) satisfying the initial conditions is given by

$$N_A(\vec{x},t) = \frac{\sigma_{eff} I(\vec{x})}{\sigma_{eff} I(\vec{x}) + k_{BA}} \left( \frac{k_{BA}}{\sigma_{eff} I(\vec{x})} + \exp\left[-\left(\sigma_{eff} I(\vec{x}) + k_{BA}\right)t\right] \right) \tag{27}$$

For $t \to \infty$ we get the steady state result

$$N_A(\vec{x}) = \frac{k_{BA}}{\sigma_{eff} I(\vec{x}) + k_{BA}} N = \frac{N}{1 + \frac{I(\vec{x})}{I_{sat}}} \quad ; \quad I_{sat} = \frac{k_{BA}}{\sigma_{eff}} \tag{28}$$

We find according to the above analysis that RESOLFT method leads to the homogeneous saturation effects described by Eq. (20).

Eqs. (23-28) describe the dynamics when the molecules are in the 'on' state A which is fluorescence activated (e.g. switched 'on' with yellow light [23]). The fluorescence becomes inhibited when the fluoroscent molecule is switched to 'off' state B, which is non-fluorescent (e.g switched 'off' by blue light [23]). This switching 'on' to fluorescent molecules and switching 'off' to nonfluorescent molecules is repeated many times. The light intensity used in the RESOLFT method can be seven orders of magnitudes less than the original STED operation. The main reduction in the laser intensity follows from the fact that the saturation parameter $I_{sat}$ in RESOLFT is much smaller relative to that used in the original STED experiments.

There are two other factors which lead to a possible reduction of light intensity in RESOLFT experiments: 1) Since this method is based on homogeneous broadening (see e.g. Eq. (22)) the saturation effect occurs for weaker light intensities relative to those used in inhomogeneous broadening for the original STED experiments (see e.g. Eq. (20)). 2) By repeating the photoswitching $N$ times, i.e., $N$ experiments, before the bleaching effect, the space uncertainty can be decreased as $\Delta x = s/\sqrt{N}$ where $\Delta x$ is the error in localization, and $s$ is the standard deviation of the point spread function (PSF).

**4. Discussion, summary and conclusions**

Along all the present article it has been shown that super-resolution measurements are related to uncertainty relations, where in one dimension such relation is given by $\Delta k_x \Delta x \geq \frac{1}{2}$, and under optimal conditions approximate equality in this relation is obtained. According to this criterion one can decrease the value of $\Delta x$ and thus increase the value of $\Delta k_x$, by which the effective wavelength, which appears in Abbe limit (Eq.1), decreases and is leading to the super-resolution effect. The optimal super-resolution effects which can be obtained in optical and biological systems have been discussed in the present paper.

There are different optical [8-15] and biological [2-7,16-28] systems in which uncertainty relations can be used to obtain super-resolution measurements. In the NSOM method super-resolution measurements are made on evanescent waves in which the wavevector in the propagation direction becomes imaginary increasing the perpendicular wavevectors components. This increase in the spatial frequencies leads to smaller effective wavelengths in the object plane by which resolution is increased. Evanecent waves are ,however, decaying in the propagation direction so that the detectors should be placed very near the object plane and such requirement can be involved with certain experimental difficulties. In recent years a new field of using a special kind of materials known as *meta materials* has been developed. The explanation for super-resolution obtained in the hyperlens which uses certain meta materials has been described. We have also related the SIM method [2,4] to the use of interference between spatial modes for getting super-resolution measurements. The common feature of these optical methods, described in Section 2, is their relation to the above uncertainty relation.

Certain theoretical methods which have been developed in relation to lasers theories [30,33] have been applied in the present paper for the interpretation of various super-resolution effects in biological systems. An example of a four-level STED system has been analyzed in analogy with a four-level laser theory [30], but the special properties of the STED laser have been inserted in this analysis. In order that the STED laser will sharpen the PSF, it needs to have a pattern with zero intensity at the laser focus and nonzero intensity at the periphery. The key to achieving super-resolution is the nonlinear dependence of the depleted population on the STED laser intensity when the saturation level is approached. If the local intensity of the STED laser is higher than a certain level all fluorescence emission is suppressed. By raising the STED laser power the saturated depletion region expands without affecting fluorescence emission at the focal point because the STED laser intensity is nearly zero at this point. The saturation effects have been analyzed in the present work by taking into account laser intensity $I(\vec{x})$ which is a function of the space coordinates $\vec{x}$. The difference in the number densities of molecules between the upper and lower STED levels, which are given as $N(\vec{x}) = n_2(\vec{x}) - n_1(\vec{x})$ in the presence of the STED laser, have been derived in relative to their difference $N_0$ in the absence of the STED laser. For homogeneous line broadening the result is given by Eq. (20), while for inhomogeneous broadening it is given by Eq. (22). One should notice that in the present derivations both stimulated emission and stimulated absorption are taken into account. The saturation parameter is given as $I_s = 1/\tau_s$ where $\tau_S$ is defined in Eq. (19). Under a certain approximation $n_2(\vec{x}) \gg n_1(\vec{x})$ so that $N(\vec{x})$ represents the number density in the upper level from which the fluorescence is obtained.

While in the original STED experiments super-resolution effects are obtained by very high power, in a variant of this method known as RESOLFT, much weaker light intensities can lead to the super-resolution effects [23] . A critical point in getting this variant method is the decrease in the value of the saturation parameter $I_S$ . Other possible factors which enable super-resolution effects in weaker light intensities have been discussed.

Localization of molecules by FPALM methods has been treated in the present paper. The localization of molecules in one experiment is based on the use of only a small number of molecules which are separated, one from another, so that there will not be a destructive interference between them. The crucial point in such method is the interference between *N* fluorescenting photons under the interaction with *a read out laser* which reduces the effective wavelength as $\lambda/\sqrt{N}$ and correspondingly reducing the PSF. The relation between this effect and lasers theories has been discussed.